\documentclass[12pt]{article}
\usepackage{amsmath,amssymb,color}
\usepackage{cite,epsf}

\makeatletter \@addtoreset{equation}{section} \makeatother

\newcommand{\be}{\begin{equation}}
\newcommand{\ee}{\end{equation}}
\newcommand{\bea}{\begin{eqnarray}}
\newcommand{\eea}{\end{eqnarray}}
\def\ie{{\it i.e.}}

\def\ft#1#2{{\textstyle{\frac{\scriptstyle #1}{\scriptstyle #2}}}}
\def\fft#1#2{\frac{#1}{#2}}

\begin{document}

\begin{flushright}
January\  2008
\end{flushright}

\vspace{10pt}

\begin{center}

{\Large\bf Lightlike Wilson loops from AdS/CFT}

\vspace{30pt}

{\large Philip C. Argyres$^1$, Mohammad Edalati$^2$,\\ [2mm] 
and Justin F. V\'azquez-Poritz$^3$}

\vspace{20pt}

{\it $^1$Department of Physics\\ University of Cincinnati, Cincinnati OH 45221-0011}\\{\tt argyres@physics.uc.edu}\\ [4mm]
{\it $^2$Department of Physics\\ University of Illinois at Urbana-Champaign, Urbana IL 61801}
\\{\tt edalati@uiuc.edu}\\ [4mm]
{\it $^3$Department of Natural Sciences\\ Baruch College, The City University of New York, New York NY 10010}\\ {\tt Justin\_Vazquez-Poritz@baruch.cuny.edu}

\vspace{20pt}

\centerline{{\bf{Abstract}}}
\end{center}

\noindent 

We investigate the lightlike limit of stationary spacelike string configurations on a large class of five-dimensional asymptotically AdS backgrounds.  Specific examples include gravity duals which incorporate finite 't Hooft coupling, curvature-squared corrections, and chemical potentials.  A universal feature of these AdS/CFT models is that the string solution with minimum action yields a lightlike Wilson loop whose leading behavior is exponentially linear, rather than quadratic, in the width of the loop.  Unless there is a compelling reason for discarding the leading saddlepoint contribution to the Wilson loop, following the proposal of Liu {\it et.\ al.}\ \cite{lrw0605} leads to zero jet-quenching parameter for all of these models.

\newpage

\section{Introduction}


The AdS/CFT correspondence \cite{agmoo9905} can be used to study certain strongly coupled gauge theory plasmas. In particular, the dynamics of open strings on a five-dimensional AdS black hole background are related to that of partons in the large $N$ and large 't Hooft coupling limit of four-dimensional ${\cal N}=4$ $SU(N)$ super Yang-Mills theory at finite temperature. Attempts to use this framework to calculate measures of the rate at which partons lose energy to the surrounding plasma, such as the friction coefficient and jet quenching parameter, have been made in \cite{lrw0605,hkkky0605,gub0605,st0605}.

Motivated by the quadratic behavior in $L$ associated with radiative energy loss by gluons in perturbative QCD, the coefficient of the $L^2$ term in the logarithm of a long lightlike Wilson loop of width $L$ was proposed as a non-perturbative definition of the jet quenching parameter in \cite{lrw0605}.  In the AdS/CFT correspondence, an open string with both endpoints on a probe brane can be used to evaluate a Wilson loop in the field theory.  In particular, the lightlike limit of a certain no-drag steadily moving string with a spacelike worldsheet was used in \cite{lrw0605,lrw0612} to compute ${\hat q}$ within this prescription.  However, in the AdS dual of ${\cal N}=4$ $SU(N)$ SYM the string configuration considered is not the solution with minimum action for the given boundary conditions, and therefore gives an exponentially suppressed contribution to the path integral \cite{argyres}.  The minimum-action solution giving the dominant saddlepoint contribution to the Wilson loop has a leading behavior that is linear in $L$, leading to the unphysical result that $\hat q = 0$.  This may either indicate that the perturbative reasoning that motivated this definition of $\hat q$ simply does not extend to strong coupling ({\it e.g.}, the mechanism for relativistic parton energy loss in the super Yang-Mills thermal bath gives a linear rather than quadratic dependence on $L$ at strong coupling), or that there is some additional physical consideration which requires that the leading saddlepoint contribution to the Wilson loop be discarded when extracting $\hat q$.

In the spirit of trying to understand the systematics of this leading saddlepoint contribution, in this paper we compute it in the AdS/CFT correspondence in a large class of five-dimensional asymptotically AdS backgrounds.  

Previously \cite{argyres}, for the ${\cal N}=4$ SYM background, we analyzed the behavior of the various saddlepoint contributions as the lightlike limit was approached in different ways.  Subsequently a number of arguments were made in \cite{lrw0612} for why the minimum-action string configuration should be discarded as unphysical.  So we will start in section 2 by reviewing the behavior of the string solutions in ${\cal N}=4$ SYM, and describing why they do not support the above-mentioned arguments.  In particular, we will further discuss the following points:
\begin{itemize}
\item
Strings which ascend above the probe brane cannot consistently be discarded on the grounds that the radius of the probe brane is a sharp UV cut-off.
\item
The dominant string does not probe shorter length scales than the thickness of the lightlike Wilson line.
\item
Due to the nature of the lightlike limit, the dynamics of the dominant string remains sensitive to the IR physics even though the entire string moves infinitely far from the black hole.
\item
The negative $L$-independent term in the action is an artifact of a regularization procedure which is ambiguous up to finite terms; it is only the difference of such terms between different string configurations which is meaningful.
\end{itemize}

In section 3, we consider spacelike strings in a large class of 5-dimensional backgrounds that are asymptotically AdS and smooth down to an event horizon. We find that in general the dominant string configuration yields a lightlike Wilson loop whose leading behavior is linear in $L$.  Then, in section 4, we review the case of a neutral AdS black hole background, and take into account the sub-leading effects of two types of corrections to this background: $\alpha^{\prime}$ corrections and curvature-squared corrections.  In section 5, we discuss spacelike strings on a three-charge AdS black hole, which corresponds to a field theory with finite chemical potentials.   Lastly, in section 6 we discuss a mass-dependent speed limit for quarks that arises for all of these backgrounds.

\section{Review of spacelike strings}

For the purpose of applying the non-perturbative definition of the jet quenching parameter given in \cite{lrw0605}, we are interested in stationary spacelike configurations for which both endpoints move along the probe brane in a direction perpendicular to their separation. As discussed in detail in \cite{argyres}, there are infinitely many of such spacelike string configurations for given endpoints, which makes it crucial to understand which branch of solutions is physically relevant. In particular, there are ``down strings" which descend below the probe brane and then turn back up, as well as ``up strings" which ascend above the probe brane and turn back down. There are also string configurations with multiple turning points that alternate above and below the probe brane. These solutions can be constructed by simply alternating segments of up and down strings. 

For spacelike worldsheets, there is a sign ambiguity in the Nambu-Goto action. Depending on the choice of sign, the string with the minimum or maximum action will dominate the path integral. The action   is proportional to the length of the string. However, since there are string configurations with arbitrarily many turns, the length of the spacelike string solutions is unbounded from above (to the extent that it does not break the probe approximation). This fixes the sign ambiguity on physical grounds and so it is the solution with minimum action that exponentially dominates the path integral. Since the shortest string must have only a single turning point, these are the solutions that we will focus on.

\begin{figure}[h]
\begin{center}
$\begin{array}{c@{\hspace{.75in}}c}
\epsfxsize=2.0in \epsffile{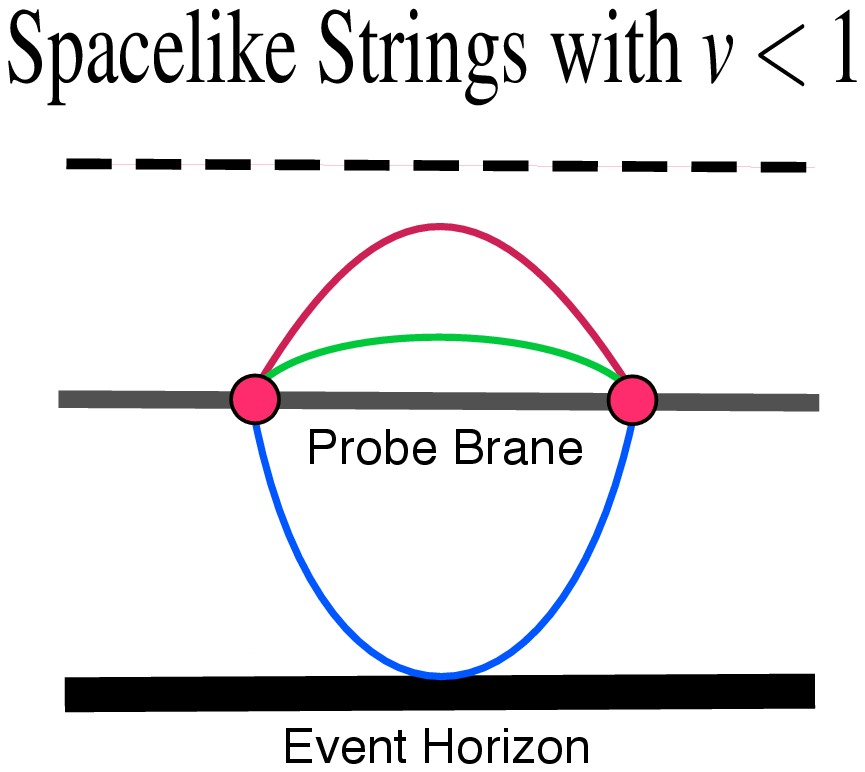} &
\epsfxsize=1.9in \epsffile{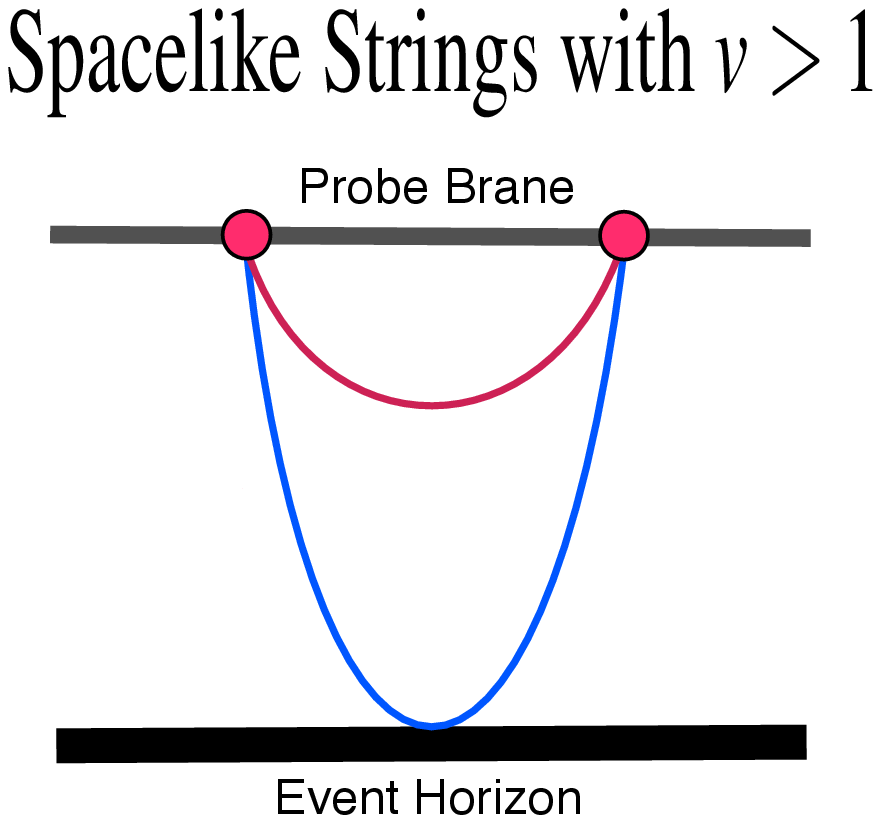}
\end{array}$
\end{center}
\caption[FIG. \arabic{figure}.]{\footnotesize{Spacelike string configurations with velocity parameter $v<1$ and $v>1$ are shown on the left and right, respectively. Spacelike strings with $v<1$ only exist below a certain radius, which is denoted by the dashed line.}}
   \label{fig3}
\end{figure}

Spacelike string configurations which have a single turning point are shown in Figure 1. The quark separation is given by the distance $L$ between the string endpoints. These strings have markedly different behavior depending on whether the worldsheet velocity parameter $v$ is greater than or less than $1$. Although spacelike configurations with $v>1$ can exist at all distances from the black hole, spacelike strings with $v<1$ can only exist below the radius $z=\sqrt{\gamma}$ (denoted by the dashed line), where $\gamma=1/\sqrt{1-v^2}$. We are using the dimensionless radial variable $z$, for which the black hole horizon is located at $z=1$.  The radius (or minimal radius, in the case of a D7-brane) of the probe brane is at $z=z_7$, which is related to the bare mass of the corresponding quarks.  For $v<1$ there are two up strings, as well as a down string with a turning point on the event horizon. On the other hand, for $v>1$ there are no up strings and two down strings---one with a turning point on the event horizon and one above it.

The relation between the velocity parameter $v$ and the proper velocity $V$ of the string endpoints generally involves $z_7$ \cite{argyres, aev0608}. For instance, on the background of a neutral AdS black hole, $V=v/\sqrt{1-z_7^{-4}}$. The physically sensible limit in which to evaluate the Wilson loop is in the lightlike limit, taking the quarks to be infinitely massive ($z_7\to\infty$) at fixed quark separation $L$. Therefore, the lightlike limit $V\to 1$ involves simultaneously taking $v\to 1$ and $z_7\to\infty$.  For other asymptotically AdS backgrounds, the timelike coordinate can always be rescaled so that the speed of light corresponds to $v=1$ in the asymptotic region.

{\it A priori}, it is not obvious that taking $v\to 1$ commutes with taking $z_7\to\infty$. Four different approaches to this limit were examined in detail in \cite{argyres}, which are labeled as (a) through (d) in Figure 2. The (a) and (b) limits apply to the $v<1$ string configurations shown in Figure 1, while the (c) and (d) limits apply to the $v>1$ strings. However, limit (a) is not interesting since it requires that $L\to 0$, in contradiction with our prescription of keeping $L$ fixed. Also, since the shorter of the two up strings with $v < 1$ (shown in green in Figure 1) does not exist with fixed $L$ in the lightlike limit, we will not consider this configuration. 

\begin{figure}[h]
\begin{center}
$\begin{array}{c@{\hspace{.75in}}c}
\epsfxsize=2.0in \epsffile{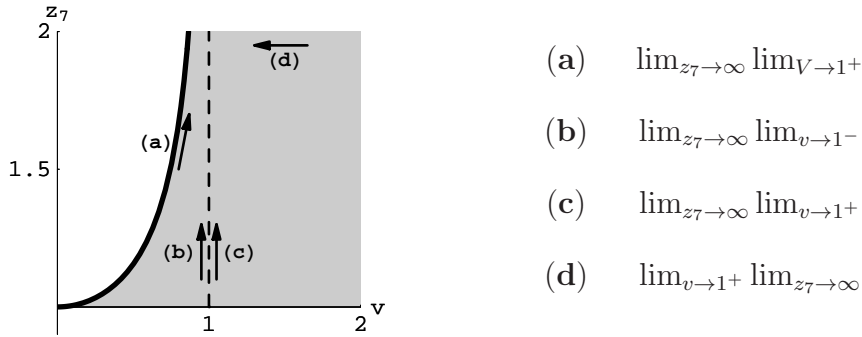} &
\begin{array}{c@{\hspace{.25in}}c} \\ [-2.1in]
{\bf (a)} & \lim_{z_7\to\infty}\lim_{V\to1^+} \\ [.45cm]
{\bf (b)} & \lim_{z_7\to\infty} \lim_{v\to1^-} \\ [.45cm]
{\bf (c)} & \lim_{z_7\to\infty} \lim_{v\to1^+} \\ [.45cm]
{\bf (d)} & \lim_{v\to1^+} \lim_{z_7\to\infty} 
\end{array} \\ [-.3cm]
\end{array}$
\end{center}
\caption[FIG. \arabic{figure}.]{\footnotesize{The shaded
   region is the set of $(v,z_7)$ for which the string worldsheet
   is spacelike and outside the horizon.  The curved boundary
   corresponds to lightlike worldsheets.  Different approaches to the
   lightlike $z_7=\infty$ limit are shown.}}
   \label{fig1}
\end{figure}

We will now discuss the behavior of the action of the remaining string configurations in the various lightlike limits. The (b) limit of the $v<1$ red string and the (c) and (d) limits of the $v>1$ red string all yield the regularized action
\be\label{redstring}
S_r(\mbox{red string})
={T\sqrt{\lambda}\over\beta}
\left(-1.31+\fft{\pi}{2} \fft{L}{\beta}\right)\,,
\ee
where $\beta$ is the inverse temperature of the black hole, $\lambda$ is the 't Hooft coupling constant and $T$ is the time interval. This is an exact expression, in the sense that no higher powers of $L$ enter. In particular, there is no $L^2$ term. As will be discussed, the constant term is from the regularization scheme. 

On the other hand, the (b) limit of the $v<1$ blue string and the (c) and (d) limits of the $v>1$ blue string all yield the regularized action
\be\label{blueS}
S_r(\mbox{blue string})
={T\sqrt{\lambda}\over\beta}
\left(0.941 \fft{L^2}{\beta^2}+{\cal O}(L^4)\right)\,.
\ee
It is reassuring that the value of the path integral does not jump discontinuously between the (b), (c) and (d) limits even though they are evaluated on qualitatively different string configurations. Although there are, in principle, many different lightlike limits intermediate between the (b), (c) and (d) limits, we take this agreement as evidence that the result is independent of how the lightlike limit is taken.

Since the red string configurations in Figure 1 have the minimum action, we will refer to these as ``short strings". Note that the short string can be an up or down string, depending on how the lightlike limit is taken.  Likewise, the blue strings will be referred to as ``long strings". 

The Wilson loop is then given by
\be
W = C_1 e^{-S_r(\mbox{\small short string})} 
+ C_2 e^{-S_r(\mbox{\small long string})} +
\cdots 
\ee
for some constants $C_i$.  In the large-$T$ (long Wilson loop) limit, this is dominated by the short string saddlepoint configuration.  A naive application of the prescription of \cite{lrw0605} that $\hat q$ is proportional to the coefficient of $L^2$ in $\ln W$ in the large-$T$ limit, would then yield the unphysical result that ${\hat q}=0$.  Thus this prescription has been modified by dropping the leading saddlepoint contribution to $W$ altogether in order to get ${\hat q}\neq0$.

The question thus arises if there is a simple physical justification for discarding the leading saddlepoint.  A number of such reasons were suggested in \cite{lrw0612}.  However, they do not seem to be supported by a detailed examination of the short string solutions, as we now discuss.

\bigskip\bigskip
\noindent\textbf{\underline{\large UV cut-off}}
\bigskip

Depending on how the lightlike limit is taken, the minimum-action string either descends down to a turning point below the radius of the probe brane or ascends up to a turning point above the probe brane. We will refer to these configurations as ``down strings" and ``up strings", respectively. If the radius of the probe brane is regarded as a sharp UV cut-off, then one might presume that the up string should be discarded, since it probes the region above the cut-off. However, in a model which treats the probe brane radius as a boundary cut-off, one does not know how to compute accurately in the AdS/CFT correspondence.  For this reason, we deal with the ${\cal N}=4$ super Yang Mills theory, for which there is no UV cut-off and the AdS/CFT correspondence is precise. 

Moreover, if one discards the up string on the premise that we will eventually have a better understanding of the AdS/CFT correspondence in the presence of a UV cut-off, then this leads to a discrepancy.  Namely, there are lightlike limits for which the dominant solution is a down string (\ie, when $v>1$), which cannot be discarded on the grounds of such a UV cut-off. The computation of the jet quenching parameter would then be ambiguous, since it depends on how the lightlike limit is taken.

\bigskip\bigskip
\noindent\textbf{\underline{\large Probing shorter length scales}}
\bigskip

As shown on the upper left portion of Figure 3, the red up string ascends infinitely far above the probe brane as $v\to 1^-$. This was taken in \cite{lrw0612} to mean that such a string probes the physics at length scales infinitely shorter than the thickness of the Wilson line. However, step 2 in version (b) of the lightlike limit is to take $z_7\to\infty$, for which the up string flattens along the probe brane. This is shown in the upper right portion of Figure 3. Thus, this string does not probe shorter length scales than the thickness of the Wilson line. Likewise, the bottom of Figure 3 shows how the second step of the (d) limit causes the red down string to flatten along the probe brane from below.

\begin{figure}[h]
\begin{center}
$\begin{array}{c@{\hspace{.75in}}c}
\epsfxsize=1.7in \epsffile{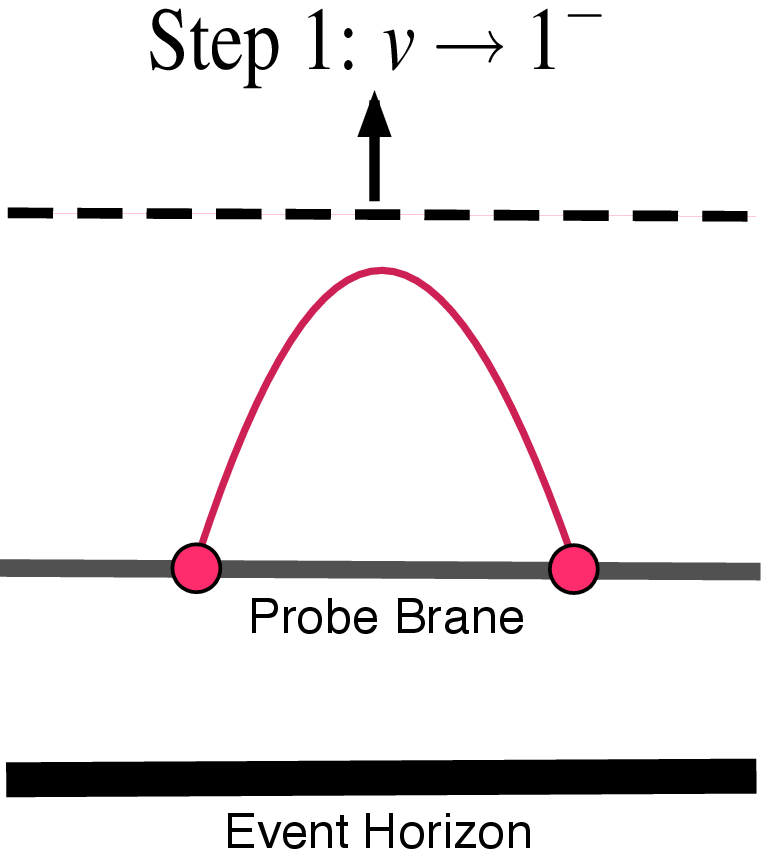} &
\epsfxsize=1.7in \epsffile{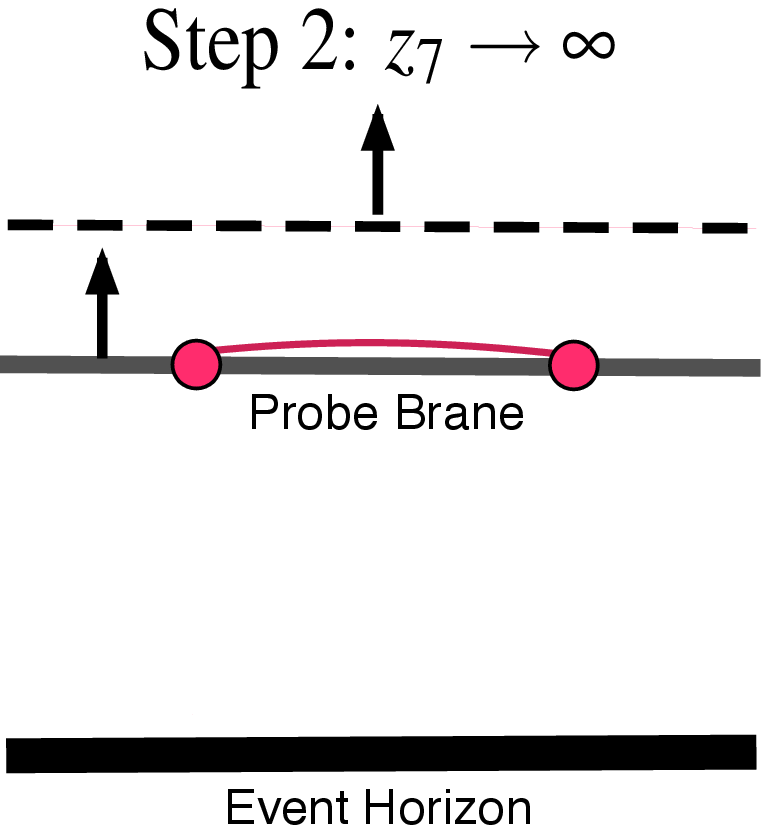}
\vspace{.1in}
\end{array}$
\line(1,0){300}\\
\vspace{.1in}
$\begin{array}{c@{\hspace{.75in}}c}
\epsfxsize=1.7in \epsffile{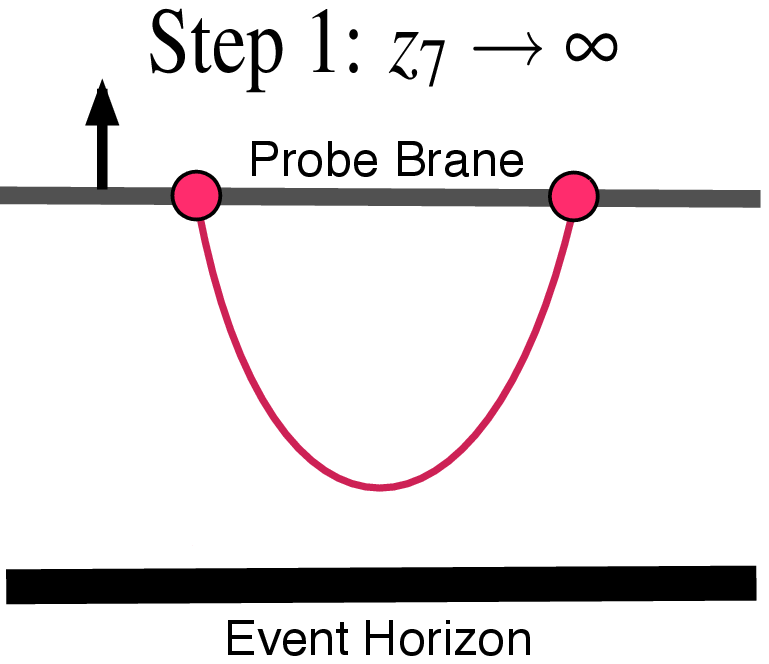} &
\epsfxsize=1.7in \epsffile{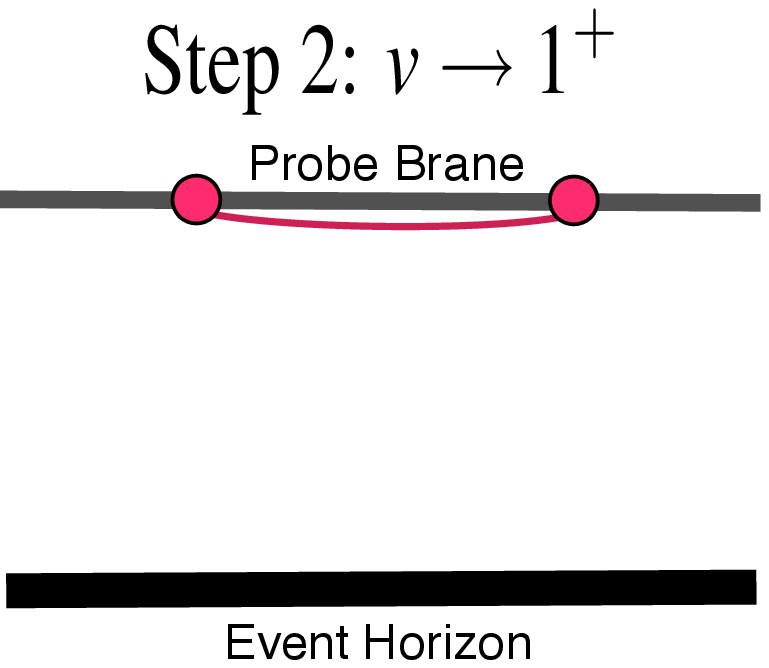}
\end{array}$
\end{center}
\caption[FIG. \arabic{figure}.]{\footnotesize{The top pictures show how an up string behaves during the two steps in version (b) of the lightlike limit. Likewise, the bottom pictures show the behavior of a down string during the two steps of version (d) of the lightlike limit. Although these two ways of taking the lightlike limit are qualitatively different, both strings approach a straight string that lies along the probe brane.}}
\end{figure}

\bigskip
\noindent\textbf{\underline{\large Sensitivity to IR physics}}
\bigskip

Regardless of how the lightlike limit is taken, the entire string configuration with minimum action moves infinitely far from the black hole. This might naturally lead one to conclude that this string is not relevant for physical observables having to do with the IR physics. To see why this is not the case, let us consider the up string. In order for the worldsheet of the up string to be spacelike, it must lie within a region bounded by the radius $z=\sqrt{\gamma}$. In this respect, this radius plays a role analogous to that of the ergosphere of a Kerr black hole. The very presence of this critical radius, for a given value of the string velocity parameter $v$, is a feature of the black hole background. Since the entire string lies within this critical radius as the lightlike limit is taken, its dynamics remain sensitive to the IR physics associated with black hole.

As can be seen in Figure 3, in the lightlike limit the up and down strings with minimum action both approach a straight string that lies along the probe brane. A hypothetical string configuration lying straight along a constant radius was briefly discussed in \cite{lrw0612}, where it was pointed out that such a string does not actually arise as a solution of the second-order equations of motion (it merely solves the first integral arising from the equations of motion) and should therefore be rejected.  We emphasize that the strings considered here are \emph{not} this straight string, even though they approach a straight string as the probe brane radius goes to infinity. Again, the fact that these strings lie within the critical radius $z=\sqrt{\gamma}$ enables them to arise as genuine spacelike solutions to the full equations of motion.

\bigskip\bigskip
\noindent\textbf{\underline{\large Negative term in action}}
\bigskip

In order to regulate the action away from the lightlike limit, we subtract the action of two straight strings which extend from the probe brane to the event horizon. For the minimum-action string, this particular regularization scheme yields a negative $L$-independent term in the action. It was claimed in \cite{lrw0612} that this leads to an unphysical result---namely, a dipole of zero size would have a nonzero photoabsorption probability. 

However, the negative term in the action is simply an artifact of this regularization procedure. In particular, throughout the literature, this subtraction is chosen in order to remove infinite constant contributions but is ambiguous up to finite terms. It has been shown in \cite{drukker} that the correct treatment of the boundary conditions together with a Legendre transform of the Nambu-Goto action (area of the worldsheet) should automatically and uniquely subtract divergent contributions from 1/2 BPS Wilson loops. It would be interesting to evaluate our (non-supersymmetric) Wilson loop using perhaps a modification of this prescription. Nevertheless, the currently used {\it ad hoc} regularization scheme suffices for our purposes of comparing the actions of two different string configurations, since this ambiguity does not affect the {\it difference} between the actions. Moreover, information regarding the energy loss of partons might be encoded within the $L$-dependent term, which does not share in this ambiguity.

\bigskip

Just because these simple criteria fail to separate the short string solutions from the long ones, it does not follow that there is not, nevertheless, some qualitative distinction between these solutions.  Evidence of such a qualitative separation between them might give a clue as to whether or why the long strings are associated to parton frictional forces while the short strings are not.  In the remainder of the paper we try to test this possibility by calculating the lightlike limits of string configurations with smallest action for a large class of asymptotically AdS black hole backgrounds.

\section{Spacelike strings on asymptotically AdS\\ backgrounds}

Consider a five-dimensional background with a metric of the form
\be
ds_5^2=-c_T^2\ dt^2+c_X^2\ (dx_1^2+dx_2^2+dx_3^2)+c_R^2\ dr^2\,,\label{metricform}
\ee
where $c_T$, $c_X$, and $c_R$ are functions only of the radial coordinate
$r$.
In the cases that we will consider, there is an event horizon located at $r_h$ and the geometry is asymptotically AdS with a radius $R$. The inverse temperature is given by
\be
\beta=\fft{4\pi c_T c_R}{(c_T^2)^{\prime}}\Big|_{r=r_h}\,.
\ee

The classical dynamics of a string in this background is described by the Nambu-Goto action
\be
S=-\fft{1}{2\pi\alpha^{\prime}}\int d\sigma\ d\tau\ \sqrt{-G}\,,\label{action}
\ee
with 
\be
G={\rm det}[g_{\mu\nu}(\partial X^{\mu}/\partial\xi^{\alpha})(\partial X^{\nu}/\partial \xi^{\beta})]\,,
\ee
where $\xi^{\alpha}=\{ \tau,\sigma\}$ and $X^{\mu}=\{ t,x_1,x_2,x_3,r\}$. 

A quark-antiquark pair with constant separation and moving with constant velocity perpendicular to the separation of the quarks can be described by the worldsheet embedding
\be
t=\tau\,,\qquad x_1=v\tau\,,\qquad x_2=\sigma\,,\qquad x_3=0\,,\qquad r=r(\sigma)\,.\label{embedding}
\ee
We take the boundary conditions
\be
0\le\tau\le T\,,\qquad -L/2\le\sigma\le L/2\,,\qquad r(\pm L/2)=r_7\,,\label{bndry}
\ee
where $r_7$ is the (minimal) radius of a probe D-brane and $r(\sigma)$ is a smooth embedding. With the embedding (\ref{embedding}) and boundary conditions (\ref{bndry}), the string action becomes
\be
S=-\fft{T}{\pi\alpha^{\prime}} \int_0^{L/2} d\sigma \sqrt{( c_T^2-v^2 c_X^2)\Big( c_X^2+c_R^2r^{\prime 2}\Big) }\,,\label{Swithsigma}
\ee
where $r^{\prime}=\partial r/\partial\sigma$. The resulting equations of motion give
\be
r^{\prime 2}=-\fft{c_X^2}{\alpha^2 c_R^2} [(c_T^2-v^2 c_X^2) c_X^2+\alpha^2 ]\,.\label{eom}
\ee
where the integration constant $\alpha^2>0$ describes spacelike string configurations. Here, we have taken the first integral of the second order equations of motion which follows from the existence of a conserved momentum in the direction along the separation of the string endpoints. 

Equation (\ref{eom}) can be integrated to give
\be
L= 2\alpha \Big| \int_{r_t}^{r_7} dr\ \fft{c_R}{c_X \sqrt{(c_T^2-v^2 c_x^2) c_X^2+\alpha^2}}\Big| \,.\label{L}
\ee
This integral expression determines $\alpha$ in terms of $L$. The absolute value is there in order to cover both $z_t<z_7$ and $z_7<z_t$. 

For the solutions of (\ref{eom}), the regulated action can be written as
\be
S_r= \fft{T \sqrt{\lambda}}{\pi R^2} \Big[ \Big| \int_{r_t}^{r_7} dr\ \fft{c_X c_R (c_T^2-v^2 c_X^2)}{\sqrt{(c_T^2-v^2 c_X^2)c_X^2+\alpha^2}}\Big| -\Big| \int_{r_h}^{r_7} dr\ c_R \sqrt{c_T^2-v^2 c_X^2}\Big| \Big]\,.\label{S}
\ee
where we have used the fact that $\alpha^{\prime}=R^2/\sqrt{\lambda}$. In order to regulate the action, we have subtracted the action of two straight strings. 

For the backgrounds that we will consider, the turning point $r_t$ is given by the root of either $(c_R)^{-2}$ or $(c_T^2-v^2 c_X^2)c_X^2+\alpha^2$. As we will see for particular examples, the first string descends all the way down to the event horizon of the black hole, even as the endpoints on a probe brane are taken to infinity. On the other hand, the second string approaches a straight string that lies along a constant radius as the probe brane is taken to infinity. Therefore, we will refer to the first string as the ``long string" and the second string as the ``short string". 

The existence of multiple strings with the same endpoints is due to the fact that multiple values of $\alpha$ correspond to the same $L$, as can be seen by analyzing (\ref{L}). In particular, there are two possibilities that correspond to small $L$, which is the relevant regime for jet quenching. Firstly, $L\rightarrow 0$ corresponds to $\alpha\rightarrow 0$. Since this holds for a finite range of integration, this applies to the long string. The other possible way of getting $L\rightarrow 0$ is for the range of integration to vanish ($r_7-r_t\rightarrow 0$) for some finite value of $\alpha$. We will see that this is indeed the case for the short string.

We will now give the general formulae for the regularized action of the long and short string configurations in the lightlike limit as $L\rightarrow 0$.

\bigskip\bigskip
\noindent\textbf{\underline{\large Long string}}
\bigskip

We will now make use of the fact that $L\rightarrow 0$ corresponds to $\alpha\rightarrow 0$ for the long string to find a general relation between $S_r$ and $L$ in this limit. For the long string, $L$ becomes
\be\label{I}
L=2\alpha I\,,\qquad I\equiv\int_{r_h}^{\infty} dr\ \fft{c_R}{c_X^2 \sqrt{c_X^2-c_T^2}}
\ee
in the lightlike limit and as $\alpha\rightarrow 0$. We are assuming that $c_X^2>c_T^2$ outside of the horizon, which tends to be the case for most black hole geometries. Likewise, the regularized action becomes
\be
S_r=\alpha^2 \fft{T\sqrt{\lambda}}{2\pi R^2} I\,.
\ee
Eliminating $\alpha$ between these two expressions gives
\be
S_r=\fft{T\sqrt{\lambda}}{8\pi R^2}\ I^{-1} L^2+{\cal O}(L^4)\,.\label{longS}
\ee
Note that this differs from the action for a general background that was given in \cite{b0605} since here it is expressed in the reference frame of the plasma rather than that of the parton. 

The fact that the regularized action vanishes in the limit $L\rightarrow 0$ can be understood by the following heuristic reasoning. As $L\rightarrow 0$, the long string becomes two coincident straight strings which extend from the black hole horizon to infinity. Since it is the action of this configuration that has been subtracted from that of the long string, the regularized action must therefore vanish.

\bigskip\bigskip
\noindent\textbf{\underline{\large Short string}}
\bigskip

For the short string, in the lightlike limit $r_t$ and $r_7$ go to infinity simultaneously. As will be demonstrated for explicit examples, in this limit $r_7-r_t\rightarrow 0$ for finite $\alpha$. Then the action reduces to that of a hypothetical straight string extended along a constant radius. Note that, while this trivial string configuration does indeed solve (\ref{eom}), it is not actually a solution of the second-order equations of motion themselves. Nevertheless, the short string, which is a genuine solution to the equations of motion, approaches this straight string configuration in the lightlike limit.

This enables us to easily evaluate the integral for the action given by (\ref{Swithsigma}) at $r=r_7$. Then the regularized action reduces to
\be\label{generalshortS}
\fft{\pi R^2}{T\sqrt{\lambda}}\ S_r=-\int_{r_h}^{\infty} dr\ c_R\sqrt{c_X^2-c_T^2}+\fft12 \Big( c_X\sqrt{c_X^2-c_T^2}\Big)\Big|_{r=r_7} L\,.
\ee

Thus, we see that the action of the short string is generally less than that of the long string. This is to be expected for the following reason. As $r_7-r_t\rightarrow 0$ and $L\rightarrow 0$, the short string has infinitesimal length and the unregularized action vanishes. Thus, in this limit $S_r\to -S_0$, where $S_0>0$ is the action of the two straight strings.  

We have assumed that the short string has a well-defined lightlike limit, for which the regularized action is not divergent. While this turns out to be the case for all asymptotically AdS backgrounds, this assumption does not necessarily carry over to backgrounds with different asymptotical behavior. For example, it has been found that none of the particular lightlike limits studied here are well-defined for a short string on ten-dimensional wrapped fivebrane duals of SQCD-like theories \cite{SQCD1}. In particular, while the unregularized action of the short string is in fact well-behaved, $S_0$ itself is divergent and so the ``regularized" action goes to negative infinity. One should therefore discard the short string solution and turn to the configuration with next lowest (and finite) action, namely the long string. However, in this case the regularized action of the long string vanishes. Thus, according to the proposed non-perturbative definition of the jet quenching parameter given in \cite{lrw0605}, ${\hat q}=0$ \cite{SQCD2}. It has been suggested that this result is associated with the non-local Little String Theory modes that are present in the UV regime. It would be interesting to see if gravitational backgrounds with $B$ fields which are dual to the large $N$ limit of non-commutative gauge theories share this property \cite{B1,B2}.

\section{Spacelike strings on a neutral AdS black hole and various corrections}

\subsection{On the neutral AdS black hole}

At finite temperature, the large $N$, large 't Hooft coupling limit of four-dimensional ${\cal N}=4$ $SU(N)$ super Yang-Mills theory is equivalent to type IIB string theory on the background of the near-horizon region of a large number $N$ of non-extremal D3-branes \cite{agmoo9905}. From the perspective of five-dimensional gauged supergravity, this is the background of a neutral AdS black hole whose Hawking temperature equals the temperature of the gauge theory. We will now apply the general formulae of the previous section to the case of a five-dimensional AdS black hole, which has the metric components
\be
c_T^2=\fft{r^2}{R^2} f\,,\qquad c_X^2=\fft{r^2}{R^2}\,,\qquad c_R^2=\fft{R^2}{r^2}f^{-1}\,,\label{AdSbh}
\ee
where
\be
f=1-\fft{r_0^4}{r^4}\,.
\ee
The radius of the event horizon $r_h=r_0$. We will use the rescaled coordinate $z=r/r_0$, so that $z_h=1$ and $z_7=r_7/r_0$. The inverse temperature is
\be
\beta=\fft{\pi R^2}{r_0}\,.\label{beta}
\ee

As can be seen by applying the general formula (\ref{eom}) to this background, the turning point of the long string is given by $z_t=1$, while the turning point of the short string is specified by $z_t^4=\gamma^2 (1-\alpha^2)$. We will now consider the lightlike limit for both string configurations.

\newpage
\noindent\textbf{\underline{\large Long string}}
\bigskip

For the case of the AdS black hole background, the general expression for the action in (\ref{longS}) reduces to (\ref{blueS}).
Note that this result is independent of the precise way in which the lightlike limit is taken. Namely, the simultaneous limits $r_7\rightarrow\infty$ and $\gamma\rightarrow\infty$, along with $\alpha\rightarrow 0$, can be applied directly to (\ref{L}) and (\ref{S}) without any ambiguity.

\bigskip\bigskip
\noindent\textbf{\underline{\large Short string}}
\bigskip

The turning point is specified by $z_t^4=\gamma^2 (1-\alpha^2)$. Since $z_t$ depends on $\gamma$, there is an ambiguity in the lightlike limit. Namely, there is a continuous family of limits for which we simultaneously take $\gamma\rightarrow\infty$ and $z_7^{-1}\equiv\epsilon\rightarrow 0$. For the AdS black hole background, we demonstrated in \cite{argyres} that three particular ways of taking the lightlike limit (the (b), (c) and (d) limits) yields the same relation (\ref{redstring}) between the regularized action and $L$, which is an indication that this relation is independent of the precise way in which the lightlike limit is taken. 

This agreement is reassuring, since it demonstrates that the path integral does not jump discontinuously between these limits even though they are evaluated on qualitatively different string configurations (namely, the (b) limit considers an string which rises above the probe brane whereas the (c) and (d) limits consider a string that descends below). When we consider the short string in other backgrounds, we will be content to restrict ourselves to the (b) limit
where we first take $v\rightarrow 1^-\ (\gamma\rightarrow\infty)$ before taking $\epsilon\rightarrow 0$.

\subsection{$\alpha^{\prime}$ corrections}

Corrections in inverse 't Hooft coupling $1/\lambda$ correspond to $\alpha^{\prime}$ corrections on the string theory side. The $\alpha^{\prime}$-corrected near extremal D3-brane has the metric \cite{alpha1,alpha2}
\be
ds_{10}^2=-c_T^2\ dt^2+c_X^2\ dx_i^2+c_R^2\ dr^2+G_{Mn}\ dx^M dx^n\,,
\ee
where $x^M=(t,x^i,r; x^n)$, $i=1,2,3$ and $n=1,\dots ,5$. We rescale the metric functions as
\be
c_T^2=\fft{r_0^2}{R^2}\ {\hat c}_T^2\,,\qquad c_X^2=\fft{r_0^2}{R^2}\ {\hat c}_X^2\,,\qquad c_R^2=\fft{R^2}{r_0^2}\ {\hat c}_R^2\,,
\ee
where
\bea
{\hat c}_T^2 &=& z^2(1-z^{-4})(1+b\ T(z)+\dots )\,,\nonumber\\
{\hat c}_X^2 &=& z^2(1+b\ X(z)+\dots )\,,\nonumber\\
{\hat c}_R^2 &=& z^{-2} (1-z^{-4})^{-1} (1+b\ R(z)+\dots )\,,\label{alphametric1}
\eea
and
\bea
T(z) &=& -75z^{-4}-\fft{1225}{16}z^{-8}+\fft{695}{16}z^{-12}\,,\nonumber\\
X(z) &=& -\fft{25}{16}z^{-8}(1+z^{-4})\,,\nonumber\\
R(z) &=& 75z^{-4}+\fft{1175}{16}z^{-8}-\fft{4585}{16}z^{-12}\,.\label{alphametric2}
\eea
The horizon is located at $z=1$, and we are using the rescaled coordinate $z=r/r_0$. The expansion parameter $b$ can be expressed in terms of the inverse 't Hooft coupling as
\be\label{blambda}
b=\fft{\zeta(3)}{8}\lambda^{-3/2}\sim .15\lambda^{-3/2}\,.
\ee

The inverse temperature $\beta$ is given by
\be
\beta =\fft{\pi R^2}{r_0}\ (1-15b)\,.\label{temp}
\ee

With the embedding (\ref{embedding}) and boundary conditions (\ref{bndry}), we have
\be
L= \Big| \fft{2\alpha R^2}{r_0} \int_{z_7}^{z_t} dz\ \fft{{\hat c}_R}{{\hat c}_X \sqrt{({\hat c}_T^2-v^2 {\hat c}_x^2) {\hat c}_X^2+\alpha^2}}\Big| \,,\label{L3}
\ee
and the regulated action can be written as
\be
S_r= \fft{T r_0\sqrt{\lambda}}{\pi R^2} \Big[ \Big| \int_{z_7}^{z_t} dz\ \fft{{\hat c}_X {\hat c}_R ({\hat c}_T^2-v^2 {\hat c}_X^2)}{\sqrt{({\hat c}_T^2-v^2 {\hat c}_X^2){\hat c}_X^2+\alpha^2}}\Big| -\Big| \int_1^{z_7} dz\ {\hat c}_R \sqrt{{\hat c}_T^2-v^2 {\hat c}_X^2}\Big| \Big]\,.\label{S3}
\ee

\bigskip
\noindent\textbf{\underline{\large Long string}}
\bigskip

The regularized action for the long string is given by \cite{charge4}
\bea
\fft{\beta S_r}{T\sqrt{\lambda}} &=& \fft{\pi^{3/2}\Gamma(3/4)}{2\Gamma(1/4)}\Big[ 1-\Big( 45-\fft{30725\ \sqrt{2}\pi}{924\ \Gamma(1/4)\Gamma(3/4)}\Big) b\Big] \fft{L^2}{\beta^2}+{\cal O}(L^4)\nonumber\\
&\approx& .941( 1-1.7652\lambda^{-3/2}+\cdots) \fft{L^2}{\beta^2}+{\cal O}(L^4)\,.
\eea
Note that this differs from the expression in \cite{charge4} since we are in the reference frame of the plasma rather than that of the parton.

\bigskip\bigskip
\noindent\textbf{\underline{\large Short string}}
\bigskip

For $z\gg 1$, we can approximate
\be
({\hat c}_T^2-v^2 {\hat c}_X^2) {\hat c}_X^2+\alpha^2\approx \fft{z^4}{\gamma^2}+\alpha^2-1-75b\,.
\ee
Thus, there is a turning point at large distance given by $z_t^4=\gamma^2 (1+75b-\alpha^2)$.
In the (b) limit, this enables us to write
\be
L=\fft{2\alpha\gamma R^2}{r_0} \int_{1/\epsilon}^{z_t} \fft{dz}{z^2 \sqrt{z_t^4-z^4}}+\cdots\,,\label{LforGB}
\ee
where there are subleading terms which will vanish as $\epsilon\rightarrow 0$. Examination of (\ref{LforGB}) shows that, $L$ remains finite if one takes $\epsilon\rightarrow 0$ in such a way that
\be
\delta^2\equiv \fft{\epsilon^2}{1+75b-\alpha^2}
\ee
remains fixed. Eliminating $\alpha$ in favor of $\delta$ in (\ref{LforGB}) and changing variables to $y=(\gamma\epsilon/\delta)^{-1/2}z$ gives
\be
L=\fft{2\alpha\gamma R^2}{r_0} \Big( \fft{\delta}{\gamma\epsilon}\Big)^{3/2} \int_{\fft{1}{\epsilon} \sqrt{\fft{\delta}{\gamma\epsilon}}}^1 \fft{dy}{y^2\sqrt{1-y^4}}\cdots \rightarrow \fft{2}{\pi}\beta (1+\ft{105}{2}b)\ \delta\,,
\ee
where we have taken the limit $\epsilon\rightarrow 0$. Also, we have used the temperature formula (\ref{temp}).

Likewise, the regulated action (\ref{S3}) can be written as
\bea
\fft{\pi R^2}{Tr_0\sqrt{\lambda}}\ S_r &=& \gamma \Big( \fft{\delta}{\gamma\epsilon}\Big)^{3/2} \int_{\ft{1}{\epsilon} \sqrt{\ft{\delta}{\gamma\epsilon}}}^1 \fft{dy}{y^2\sqrt{1-y^4}}\Big( 1+75b-\fft{\epsilon^2}{\delta^2}y^4+\cdots\Big)\nonumber\\ &-& \int_1^{1/\epsilon}\fft{dz}{\sqrt{z^4-1}}\Big( 1+\ft12 b(75+75z^{-4}-\ft{385}{8}z^{-8}-\ft{1945}{8}z^{-12})\Big)\,.\nonumber
\eea
Evaluating these integrals in the (b) limit and writing the action in terms of $L$ gives
\be
\fft{\beta S_r}{T\sqrt{\lambda}}=c+\fft{\pi}{2\beta}(1+1.13\lambda^{-3/2}+\cdots)L\,,
\ee
where
\be
c=-\fft{\sqrt{\pi}\Gamma[\ft54]}{\Gamma[\ft34]}\Big( 1+\fft{128845}{7392}b+\cdots\Big)\approx -1.31(1+2.61\lambda^{-3/2}+\cdots )\,,
\ee
and we have used (\ref{blambda}). This is in agreement with the general formula (\ref{generalshortS}).

\subsection{Curvature-squared corrections}

\subsubsection{General corrections}

The five-dimensional action describing general curvature-squared corrections is given by
\be
S=\int d^5x \sqrt{-g} \Big[ \fft{{\cal R}}{2\kappa}+\fft{6}{\kappa R^2}+c_1\ {\cal R}^2+c_2\ {\cal R}_{\mu\nu} {\cal R}^{\mu\nu}+c_3\ {\cal R}_{\mu\nu\rho\sigma} {\cal R}^{\mu\nu\rho\sigma}\Big] \,.
\ee
where $c_i$ are arbitrary (but small) coefficients. The metric components for a five-dimensional AdS black brane solution with the sub-leading curvature-squared corrections is given by \cite{R21}
\be
c_T^2=\fft{r^2}{R^2}fk\,,\qquad c_X^2=\fft{r^2}{R^2}\,,\qquad c_R^2=\fft{R^2}{r^2}f^{-1}\,,\label{R2metric1}
\ee
where
\be
f=1-\fft{r_0^4}{r^4}+b_1+b_2 \fft{r_0^8}{r^8}\,,\label{R2metric2}
\ee
and
\bea\label{bc}
b_1 &=& \fft{4\kappa}{3R^2}\ [2(5c_1+c_2)+c_3]\,,\nonumber\\
b_2 &=& \fft{4\kappa}{R^2}\ c_3\,.
\eea
We have included the scaling factor $k=1/(1+b_1)$ for time in order for the speed of light in the boundary theory to be unity. 

The formula for the inverse temperature of the black brane with the subleading corrections is given by
\be
\beta=\fft{\pi R^2}{r_0}\Big( 1-\fft14 b_1+\fft54 b_2\Big) \,.\label{betaR2}
\ee

It has been shown that the conjectured lower bound of $1/4\pi$ on the viscosity-to-entropy ratio \cite{bound1,bound2} can be violated by the curvature-squared corrections \cite{R21,R22}. The new ratio is given by
\be\label{viscosity}
\fft{\eta}{s}=\fft{1}{4\pi}\Big[ 1-\fft{16c_3\kappa}{R^2}\Big]\,,
\ee
which violates the conjectured bound for theories with $c_3>0$. Note that while $c_1$ and $c_2$ affect the viscosity $\eta$ and the entropy $s$ separately but not their ratio.

With the embedding (\ref{embedding}) and boundary conditions (\ref{bndry}),
\be
L=\Big| \fft{2\alpha R^2}{r_0} \int_{z_t}^{z_7} \fft{dz}{z^2 \sqrt{f[z^4(kf-v^2)+\alpha^2]}}\Big| \,,\label{LforR2}
\ee
where $z=r/r_0$. With the rescaled coordinate $z$,
\be
{\tilde f}\equiv kf=1-\fft{k}{z^4}+\fft{kb_2}{z^8}\,.
\ee
The black hole horizon $z_h$ is given by the largest root of ${\tilde f}$, which is
\be
z_h^4=\fft12 (k+\sqrt{k(k-4b_2)})\,.
\ee
The turning point $z_t$ is either at the black hole horizon $z_h$ or else at $z_+$, where
\be
z_{\pm}^4=\fft{\gamma^2}{2} \Big(k-\alpha^2\pm\sqrt{(k-\alpha^2)^2-4kb_2 \gamma^{-2}}\Big)\,,
\ee
and $\gamma=1/\sqrt{1-v^2}$.

The regulated action can be written as
\be
S_r=\fft{Tr_0\sqrt{k\lambda}}{\pi R^2}\Big[ \Big| \int_{z_t}^{z_7} dz\ \fft{({\tilde f}-v^2)z^2}{\sqrt{{\tilde f}[z^4({\tilde f}-v^2)+\alpha^2]}}\Big| -\Big| \int_{z_h}^{z_7} dz \sqrt{\fft{{\tilde f}-v^2}{{\tilde f}}}\Big| \Big]\,.\label{SforR2}
\ee

\bigskip
\noindent\textbf{\underline{\large Long string}}
\bigskip

Since regularized action for the long string configuration has not been considered for this background elsewhere, we will show some of the details. For metric components given by (\ref{R2metric1}) and (\ref{R2metric2}), the general expression (\ref{I}) for the integral $I$ becomes
\be
I=\int_{r_h}^{\infty}dr\ \fft{R^4}{r^4\sqrt{f}\sqrt{1-kf}}\,.
\ee
Using the variable $y\equiv r/r_h$, $I$ can be expanded for small $b_1$ and $b_2$ as
\bea
I &=& \Big( 1+\fft14 b_1+\fft14 b_2+\cdots\Big) \fft{R^4}{r_0^3}\int_1^{\infty} \fft{dy}{\sqrt{y^4-1}}\Big(1+\fft{b_2}{y^4}+\cdots\Big)\nonumber\\
&=& \fft{\sqrt{\pi}\Gamma(1/4)}{4\Gamma(3/4)}\fft{R^4}{r_0^3}\Big( 1+\fft14b_1+\fft{7}{12}b_2+\cdots\Big)\,.
\eea
Using (\ref{longS}) and (\ref{betaR2}), we find that the regularized action is given by
\bea
\fft{\beta S_r}{T\sqrt{\lambda}} &=& \fft{\pi^{3/2}\Gamma(3/4)}{2\Gamma(1/4)}\Big( 1-b_1+\fft{19}{6}b_2\Big) \fft{L^2}{\beta^2}+{\cal O}(L^4)\nonumber\\
&\approx& .941\Big[ 1-\fft{8\kappa}{3R^2}\Big( 5c_1+c_2-\fft{17}{4} c_3\Big)\Big] \fft{L^2}{\beta^2}+{\cal O}(L^4)\,.
\eea

\bigskip
\noindent\textbf{\underline{\large Short string}}
\bigskip

We will now consider the (b) limit. Since in this limit we first take $v\rightarrow 1^-\ (\gamma\rightarrow\infty)$ before taking $\epsilon\equiv z_7^{-1}\rightarrow 0$, (\ref{LforR2}) becomes
\be
L=\fft{2\alpha\gamma \sqrt{k} R^2}{r_0} \int_{1/\epsilon}^{z_+} \fft{dz}{\sqrt{{\tilde f}(z_+^4-z^4)(z^4-z_-^4)}}\,.
\ee
In order for $L$ to remain fixed in this limit, we must keep
\be
\delta^2\equiv \fft{2\epsilon^2}{k-\alpha^2+\sqrt{(k-\alpha^2)^2-4kb_2\gamma^{-2}}}
\ee
fixed. Upon taking $\gamma\rightarrow\infty$, this reduces to 
\be
\delta^2=\fft{\epsilon^2}{k-\alpha^2}\,.
\ee

As $\gamma\rightarrow\infty$, $z_+^4\rightarrow \gamma^2\epsilon^2/\delta^2$ and $z_-^4\rightarrow kb_2\delta^2/\epsilon^2$. Note that $1\ll z_-\ll \epsilon^{-1}$ as $\epsilon\rightarrow 0$. This enables us to expand  the $1/\sqrt{{\tilde f}}$ and $1/\sqrt{z^4-z_-^4}$ factors for large $z$. Changing variables to $y=(\gamma\epsilon/\delta)^{-1/2} z$, $L$ can be expressed as a series of Hypergeometric integrals. After taking the limits $\gamma\rightarrow\infty$ and $\epsilon\rightarrow 0$ while keeping $\delta$ fixed, we find that
\be
L=\fft{2kR^2}{r_0}\delta=\fft{2}{\pi} [1+\ft54(b_2-b_1)+\cdots]\delta\,,
\ee
where we have used the corrected formula for the inverse temperature (\ref{betaR2}).

Likewise, the regulated action (\ref{SforR2}) can be written as
\be
S_r=\fft{Tr_0\sqrt{k\lambda}}{\pi\gamma R^2}\Big[ \int_{z_7}^{z_+} dz\ \fft{\gamma^2 k(z^4-b_2)-z^8}{z^4\sqrt{{\tilde f}(z_+^4-z^4)(z^4-z_-^4)}} -\int_{z_h}^{z_7} dz \sqrt{\fft{k\gamma^2(z^4-b_2)-z^8}{z^8+k(b_2-z^4)}}\Big]\,.
\ee
We expand $1/\sqrt{{\tilde f}}$ and $1/\sqrt{z^4-z_-^4}$ in the first integral for large $z$. We also expand both integrals for small $b_2$ keeping only the linear correction term. The result is that
\be
\fft{\beta S_r}{T\sqrt{\lambda}}=c+\Big( 1-\fft74 b_1+\fft54 b_2+\cdots\Big) \delta\,,
\ee
where
\bea
c &=& -\fft{\sqrt{\pi}\ \Gamma(5/4)}{\Gamma(3/4)}\ \Big( 1-b_1+\fft32 b_2+\cdots\Big)\nonumber\\
&\approx& -.131\Big( 1-\fft{14\kappa}{R^2}c_3\Big)\,,
\eea
where we have used (\ref{bc}). The action can be written in terms of $L$ as
\be
\fft{\beta S_r}{T\sqrt{\lambda}}=c+\fft{\pi}{2}\Big( 1-\fft{2\kappa}{3R^2} (20c_1+4c_2-13c_3)+\cdots\Big) \fft{L}{\beta}\,,
\ee
which agrees with (\ref{generalshortS}).

\subsubsection{Gauss-Bonnet gravity}

The Gauss-Bonnet combination of curvature-squared terms corresponds to setting $b_1=b_2\equiv b$. In this case, there is an exact black brane solution whose metric coefficients have the form (\ref{R2metric1}), where $f$ and $k$ are now given by \cite{gb1,gb2}
\be
f=\fft{1}{2b}\Big[ 1-\sqrt{1-4b\Big( 1-\fft{r_0^4}{r^4}\Big) }\Big]\,,\qquad k=\fft12 (1+\sqrt{1-4b})\,.
\ee
As before, we have chosen $k$ so that the boundary speed of light is unity. We assume that $b\le 1/4$, since beyond this point there is no vacuum AdS solution. 

The inverse temperature is
\be
\beta=\fft{\pi R^2}{\sqrt{k}r_0}\,.
\ee
We will now present the regularized action for the long and short strings in this background.

\bigskip\bigskip
\noindent\textbf{\underline{\large Long string}}
\bigskip

From (\ref{I}) and (\ref{longS}), we find that
\be
\fft{\beta S_r}{T\sqrt{\lambda}}=F(b)\ \fft{L^2}{\beta^2}+{\cal O}(L^4)\,,
\ee
where
\be
F(b)=\fft{\pi^2}{8\sqrt{2bk^3}}\Big[ \int_1^{\infty} \fft{dz}{z^4\sqrt{h(1-\fft{k}{2b}h)}}\Big]^{-1}\,,\qquad h\equiv 1-\sqrt{1-4b(1-z^{-4})}\,.
\ee
\begin{figure}[t]
   \epsfxsize=3.0in \centerline{\epsffile{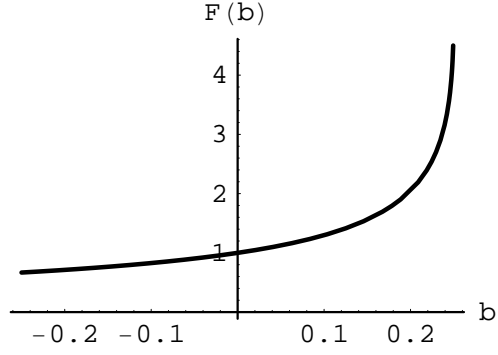}}
   \caption[FIG. \arabic{figure}.]{\footnotesize{$F(b)$ versus $b$ for the long string in the case of Gauss-Bonnet gravity.}}
\label{fig9}
\end{figure}

We numerically integrated to solve for $F(b)$, the result of which is shown in Figure 4. Note that if there are no curvature-square corrections then $F(0)=0.941$. $F(b)$ has exponential behavior up to the value $F=5.244$ at $b=1/4$. If the short string could be discarded and the jet quenching parameter read off from the $L^2$ term of the regularized action, then we would find that ${\hat q}$ is enhanced due to the Gauss-Bonnet corrections with positive $b$, while ${\hat q}$ decreases with negative $b$. However, as previously discussed, we do not know of a compelling reason to discard the short string.

\bigskip\bigskip
\noindent\textbf{\underline{\large Short string}}
\bigskip

\begin{figure}[t]
   \epsfxsize=3.0in \centerline{\epsffile{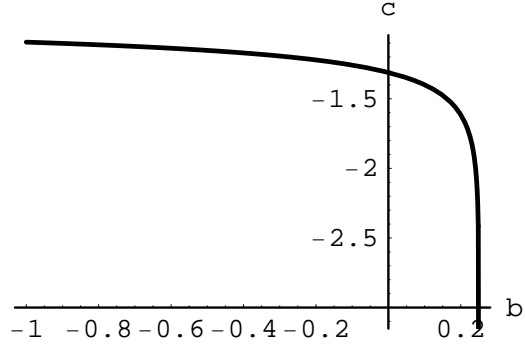}}
   \caption[FIG. \arabic{figure}.]{\footnotesize{$c$ as a function of $b$ for the short string in the case of Gauss-Bonnet gravity.}}
\label{fig10}
\end{figure}

In the lightlike limit, the regulated action of the short string can be written as
\be
\fft{\beta S_r}{T\sqrt{\lambda}}=c+\fft{\pi}{2} F(b) \fft{L}{\beta}\,,
\ee
where $F(b)$ can be solved exactly as
\be
F(b)=\sqrt{\fft{2}{\sqrt{1-4b}(1+\sqrt{1-4b})}}\,,
\ee
and
\be
c=-\int_1^{\infty} dz\ \sqrt{\fft{2b}{k(1-\sqrt{1-4b(1-z^{-4})})}-1}\,.
\ee

We numerically integrated to solve for $c$ as a function of $b$, the result of which is presented in Figure 5. Note that $c$ goes to its maximum value of $-0.826$ as $b\to-\infty$. Since $c$ is always negative, the action of the short string is less than that of the long string for all values $b$.

\section{Including chemical potentials}

A five-dimensional three-charge AdS black hole has the metric components \cite{behrndt}
\be
c_T^2={\cal H}^{-2/3}f\,,\qquad c_X^2=\fft{r^2}{R^2}\ {\cal H}^{1/3}\,,\qquad c_R^2={\cal H}^{1/3}f^{-1}\,,\label{qmetric1}
\ee
where
\be
f=\fft{r^2}{R^2}\Big( {\cal H}-\fft{r_0^4}{r^4}\Big)\,,\qquad {\cal H}=H_1H_2H_3\,,\qquad H_i=1+\fft{q_i}{r^2}\,,\label{qmetric2}
\ee
and $i=1,2,3$. 

In order to avoid a naked singularity, $q_i\ge 0$. The radius of the horizon $r_h$ is given by the largest root of $f$. For the case of vanishing $q_3$, this is given by
\be
r_h^2=\fft12 \Big( \sqrt{4r_0^4+(q_1-q_2)^2}-(q_1+q_2)\Big)\,.
\ee
For vanishing $q_2$, a regular horizon is guaranteed for all $r_0>0$. On the other hand, for two-charge case it is required that $q_1 q_2<r_0^4$ in order to have a regular horizon. A similar requirement is needed for the more complicated case of three charges.

This gravity background is dual to super Yang-Mills theory with finite temperature and finite chemical potential for the $U(1)$ R-charges. The inverse temperature $\beta$ is given by
\be
\beta =\fft{2\pi r_h^2 R^2 \prod_i \sqrt{r_h^2+q_i}}{2r_h^6+r_h^4 \sum_i q_i-\prod_i q_i}\,.\label{betachem}
\ee
The density of physical charge and chemical potentials are given by
\be
\rho_i=\fft{\sqrt{2q_i}N^2}{8\pi^2 R^6r_h}\prod_i \sqrt{r_h^2+q_i}\,,
\ee
\be
\phi_i=\fft{\sqrt{2q_i} \prod_j \sqrt{r_h^2+q_j}}{r_h R^2 (r_h^2+q_i)}\,,
\ee
respectively. We should express $q_i$ in terms of $\beta$ and $\rho_i$ for the canonical ensemble and in terms of $\beta$ and $\phi_i$ for the grand canonical ensemble.

For the string embedding given by (\ref{embedding}) with the boundary conditions (\ref{bndry}), 
\be
L=\Big| \fft{2R^2\alpha \gamma}{r_0}\ \int_{z_t}^{z_7} \fft{dz}{\sqrt{(z^4 {\cal H}-1)[{\cal H}^{-1/3}(\gamma^2-z^4 {\cal H})-\gamma^2 \alpha^2]}}\Big|\,,
\ee
where $z=r/r_0$.

The regulated action can be written as
\bea
S_r &=& \fft{T\sqrt{\lambda} r_0}{\gamma \pi R^2} \Big[ \Big| \int_{z_t}^{z_7} dz\ {\cal H}^{-1/3} \fft{\gamma^2-z^4{\cal H}}{\sqrt{(z^4{\cal H}-1)[{\cal H}^{-1/3}(\gamma^2-z^4 {\cal H})-\gamma^2 \alpha^2]}}\Big|\nonumber\\ &-& \Big| \int_{z_t}^{z_7} dz\ {\cal H}^{-1/6} \sqrt{\fft{\gamma^2-z^4{\cal H}}{(z^4{\cal H}-1)}}\Big|\Big]\,.
\eea

\bigskip
\noindent\textbf{\underline{\large Long string}}
\bigskip

The action for the long string in this background has already been extensively studied in \cite{charge1,charge2,charge3,charge4}. As an example of some of the expressions that have been obtained, we consider the case of a single charge $q_1\neq 0$, $q_2=q_3=0$ in the limit $q_1\ll r_0^2$. For the canonical ensemble, the regularized action for the long string can be expressed as
\be
\fft{\beta S_r}{T\sqrt{\lambda}}\approx 0.941\Big( 1+2.04\xi^2-11.35\xi^4+96.46\xi^6+\cdots\Big)\fft{L^2}{\beta^2}+{\cal O}(L^4)\,,
\ee
where $\rho_1\equiv\rho$, $\rho_2=\rho_3=0$ and $\xi\equiv\rho\beta^3/N^2\ll 1$. For the grand canonical ensemble,
\be
\fft{\beta S_r}{T\sqrt{\lambda}}\approx 0.941\Big( 1+0.03\zeta^2+0.0005\zeta^4+0.000008\zeta^6+\cdots\Big)\fft{L^2}{\beta^2}+{\cal O}(L^4)\,,
\ee
where $\phi_1\equiv\phi$, $\phi_2=\phi_3=0$ and $\zeta\equiv\mu\beta\ll 1$. Similar expressions have obtained for other cases, such as two equal charges $q_1=q_2$, $q_3=0$.

\newpage
\noindent\textbf{\underline{\large Short string}}
\bigskip

For large $z$, we can approximate
\be
{\cal H}^{-1/3}(\gamma^2-z^4 {\cal H})-\gamma^2 \alpha^2\approx \gamma^2(1-\alpha^2)-z^4\,.
\ee
Thus, there is a turning point at large distance given by $z_t^4=\gamma^2(1-\alpha^2)$. 
$L$ remains finite if one takes $\epsilon\rightarrow 0$ in such a way that $\delta^2\equiv \epsilon^2/(1-\alpha^2)$ remains fixed. In the (b) limit, we can express $L$ as
\be
L=\fft{2\alpha\gamma R^2}{r_0} \int_{1/\epsilon}^{z_t} \fft{dz}{z^2\sqrt{z_t^4-z^4}}+\cdots\rightarrow \fft{2R^2}{r_0}\delta\,.
\ee
Likewise, the regulated action is
\be
S_r = \fft{T\sqrt{\lambda} r_0}{\gamma \pi R^2} \Big[ \int_{1/\epsilon}^{z_t} dz\ \fft{\gamma^2-z^4}{z^2 \sqrt{z_t^4-z^4}}+\cdots - \int_{z_h}^{1/\epsilon} dz\ {\cal H}^{-1/6} \sqrt{\fft{\gamma^2-z^4{\cal H}}{z^4{\cal H}-1}}\Big]\,,
\ee
Taking the (b) limit, we find
\be
\fft{\pi R^2}{T\sqrt{\lambda}r_0}S_r=c+\fft{r_0}{2R^2} L\,,
\ee
where
\be
c=-\int_{z_h}^{\infty} dz\ \fft{{\cal H}^{-1/6}}{\sqrt{z^4{\cal H}-1}}\,.
\ee

We solved for the constant $c$ numerically as a function of $q_i$. The results for a single chemical potential ($q_1=q, q_2=q_3=0$), two equal chemical potentials ($q_1=q_2=q, q_3=0$) and three equal chemical potentials ($q_1=q_2=q_3=q$) are shown in Figure 1. The domain of the plot is determined by the region of thermodynamic stability:
\be
2r_h^6-r_h^4 \sum_i q_i-\prod q_i>0\,,
\ee
which means that $q<\ft{2}{\sqrt{3}}r_0^2$ for a single chemical potential, $q<\ft12 r_0^2$ for two equal chemical potentials and $q<.296 r_0^2$ for three equal chemical potentials.
\begin{figure}[t]
   \epsfxsize=3.5in \centerline{\epsffile{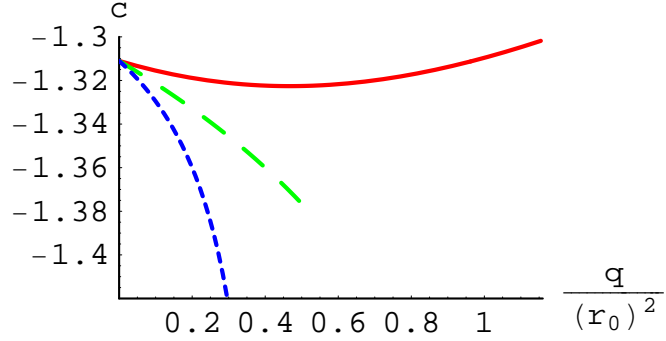}}
   \caption[FIG. \arabic{figure}.]{\footnotesize{$c$ as a function of $q_i$. For the solid red line, $q_1=q, q_2=q_3=0$. For the dashed green line, $q_1=q_2=q, q_3=0$. For the dotted blue line, $q_1=q_2=q_3=q$.}}
\label{fig2}
\end{figure}

Trading $r_0$ for $\beta$, one can write
\be
\fft{\beta S_r}{T\sqrt{\lambda}}=\sqrt{K} c+\fft{\pi}{2} K \fft{L}{\beta}\,,
\ee
where $K$ is a factor due to the presence of chemical potentials. We will now consider some simple cases. For a single chemical potential ($q_1\equiv q, q_2=q_3=0$),
\be
K=\fft{1}{2}\sqrt{\Big( 1+\sqrt{1+\fft{2q\beta^2}{\pi^2 R^4}}\Big)^2-\fft{q^2\beta^4}{\pi^4 R^8}}\,.
\ee
For two non-vanishing and equal chemical potentials ($q_1=q_2\equiv q, q_3=0$),
\be
K=1+\fft{q\beta^2}{\pi^2 R^4}\,.
\ee
In the limit $q_i\ll r_0^2$, we have
\be
K=1+\fft{\beta^2 \sum_i q_i}{2\pi^2 R^4}+\cdots\,.
\ee
Writing $q_i$ in terms of $\beta$ and $\phi_i$ or $\rho_i$ generally gives rather complicated expressions. However, this can be done rather simply for $q_i\ll r_0^2$. In the grand canonical ensemble, this limit corresponds to $\phi_i \beta\ll 1$, for which we can write
\be
K=1+\fft{\beta^2\sum_i\phi_i^2}{4\pi^2}+\cdots\,.
\ee
In the canonical ensemble, this limit corresponds to $\rho_i\beta^3/N^2\ll 1$ and
\be
K=1+\fft{16\beta^6\sum_i\rho_i^2}{\pi^2 N^4}+\cdots\,.
\ee

\section{A note on the speed limit for quarks}

Thus far, we have considered spacelike strings. We will now comment on a feature of timelike strings which comes about from the fact that the proper velocity $V$ of the endpoints differs from the worldsheet velocity parameter $v$ \cite{argyres, aev0608}. For strings moving in a background described by a metric of the form (\ref{metricform}), we have
\be
V=\fft{c_X}{c_T}v\,.
\ee
Thus, in order for a moving quark to be described by the endpoint of a timelike string, we must have $V<1$, which implies that
\be
v< v_{max}=\fft{c_T}{c_X}\Big|_{z=z_7}\,,
\ee
where $z=r/r_0$ and the radius of the endpoint is $z_7$.

For the AdS black hole with the metric components given by (\ref{AdSbh}), 
\be
v_{max}=\sqrt{1-z_7^{-4}}\,.
\ee
Recall that the probe brane radius $z_7$ is related to the bare mass of the quarks. This has recently been discussed extensively for the case of mesons in \cite{speed3}. Since this speed limit arises as a general feature of the background geometry, it should presumably also apply to single quarks and possibly even gluons.

Including the leading corrections due to finite 't Hooft coupling $\lambda$, the metric components are now given by (\ref{alphametric1}) and (\ref{alphametric2}). Then we have
\be
v_{max} =\sqrt{1-z_7^{-4}}\ \Big( 1-5.63\lambda^{-3/2}\ z_7^{-4} (1+z_7^{-4}-\fft35 z_7^{-8})\Big)\,.
\ee
Note that this correction renders $v_{max}$ smaller. 

Due to the leading ${\cal R}^2$ type corrections, the metric components are given by (\ref{R2metric1}) and (\ref{R2metric2}). Then
\be
v_{max}= \sqrt{1-z_7^{-4}}\Big( 1+\fft{b_1+b_2 z_7^{-4}}{2(z_7^4-1)}\Big) \,,
\ee
where $b_1$ and $b_2$ are given by (\ref{bc}). These corrections render $v_{max}$ larger. 

In order to include the effect of chemical potentials in the theory, we use the metric components for an AdS black hole with three $U(1)$ R-charges, which are given by (\ref{qmetric1}) and (\ref{qmetric2}). This results in
\be
v_{max}= \sqrt{1-z_7^2 \prod_i (z_7^2+q_i/r_0^2)^{-1}}\,.
\ee
Note that $v_{max}$ increases due to the charge parameters $q_i$. In the grand canonical ensemble and for $\phi_i\beta\ll 1$, this can be written as
\be
v_{max}= \sqrt{1-z_7^2 \prod_i \Big( z_7^2+\fft{\phi_i^2\beta^2}{2\pi^2}\Big)^{-1}}\,,
\ee
while in the canonical ensemble and for $\rho_i\beta^3/N^2\ll 1$, this can be expressed as
\be
v_{max}= \sqrt{1-z_7^2 \prod_i \Big( z_7^2+\fft{32}{\pi^2}\Big( \fft{\rho_i\beta^3}{N^2}\Big)^2\Big) ^{-1}}\,.
\ee

\section*{Acknowledgments}
P.C.A. is supported in part by DOE grant FG02-84-ER40153. M.E. is supported by DOE grant DE-FG02-91ER40709.

\end{document}